\documentclass{article}
\usepackage{etoolbox}

\usepackage[T1]{fontenc}
\usepackage{microtype}
\usepackage[american]{babel}

\usepackage{upgreek}
\usepackage{paralist}
\usepackage{enumitem}
\usepackage{stmaryrd}
\usepackage{graphicx}
\usepackage{xspace,xcolor}
\usepackage{amsmath,amsfonts,amssymb,mathtools}

\usepackage{bm} 
\usepackage{bbm} 

\usepackage{rsfso}
\usepackage{scalerel}
\usepackage{hyperref}
\usepackage{mathpartir}
\usepackage{prftree}
\usepackage{bussproofs}
\usepackage{lineno}

\usepackage{ellipsis}
\usepackage{placeins} 

\usepackage{tikz}
\usepackage{tikz-cd}
\usetikzlibrary{arrows}
\usetikzlibrary{decorations.pathmorphing,shapes}

\makeatletter
\g@addto@macro\normalsize{%
  \setlength\abovedisplayskip{5pt}
  \setlength\belowdisplayskip{5pt}
  \setlength\abovedisplayshortskip{4pt}
  \setlength\belowdisplayshortskip{4pt}
}
\makeatother

\newcommand{\s}{\sigma}

\newcommand{\G}{\Gamma}
\newcommand{\D}{\Delta}
\newcommand{\state}{\sigma}
\newcommand{\mstate}{\mathfrak{s}}
\newcommand{\memstate}{\mu}
\newcommand{\statevector}{\boldsymbol{\sigma}}

\DeclarePairedDelimiter\sem{\llbracket}{\rrbracket}
\DeclarePairedDelimiter\abs{\lvert{}}{\rvert}

\newcommand{\step}[2]{#1[#2]}
\newcommand{\source}[1]{\partial_0 #1}
\newcommand{\target}[1]{\partial_1 #1}
\newcommand{\sourcecode}[1]{\overline{\partial}_0 #1}

\newcommand{\length}[1]{\mathsf{len}(#1)}

\newcommand{\State}{\mathbf{State}}
\newcommand{\MStates}{\mathbf{MState}}
\newcommand{\SStates}{\mathbf{SState}}

\newcommand{\LogStates}{{\mathbf{LState}}}
\newcommand{\LStates}{\LogStates}
\newcommand{\Code}{\mathbf{Code}}
\newcommand{\Instr}{\mathbf{Instr}}
\newcommand{\RVar}{\mathbf{LockName}}

\newcommand{\Var}{\mathbf{Var}}
\newcommand{\Loc}{\mathbf{Loc}}

\newcommand{\Val}{\mathbf{Val}}
\newcommand{\Traces}{\mathbf{Traces}}
\newcommand{\Trans}{\mathbf{Trans}}
\newcommand{\pullback}[1]{#1^{\ast}}

\newcommand{\returntraces}[1]{\big|\hspace{.08em}#1\hspace{.08em}\big|}
\newcommand{\sequential}{;}

\newcommand{\Nat}{\mathbb{N}}
\newcommand{\Perm}{\mathbf{Perm}}

\newcommand{\Shuffles}[2]{\mathbf{Shuffles}(#1,#2)}

\newcommand{\strat}[2]{\mathtt{strat}({#1},{#2})}

\DeclareMathOperator{\dom}{dom}
\DeclareMathOperator{\domC}{dom_C}
\DeclareMathOperator{\domF}{dom_F}

\DeclareMathOperator{\hdom}{hdom}
\DeclareMathOperator{\vdom}{vdom}

\DeclareMathOperator{\lock}{lock}
\DeclareMathOperator{\True}{\mathbf{true}}
\DeclareMathOperator{\False}{\mathbf{false}}

\newcommand{\kw}[1]{\mathbb{\mathtt{#1}}}
\newcommand{\kwhile}{\kw{while}}
\newcommand{\kif}{\kw{if}}
\newcommand{\kthen}{\kw{then}}
\newcommand{\kelse}{\kw{else}}
\newcommand{\kalloc}{\kw{alloc}}
\newcommand{\kdispose}{\kw{dispose}}
\newcommand{\kwith}{\kw{with}}
\newcommand{\kwhen}{\kw{when}}
\newcommand{\kresource}{\kw{resource}}
\newcommand{\knop}{\kw{nop}}
\newcommand{\kskip}{\kw{skip}}
\newcommand{\kdo}{\kw{do}}
\newcommand{\ifte}[3]{\kif\: #1\: \kthen\: #2\: \kelse\: #3}
\newcommand{\while}[2]{\kwhile\: #1 \: \kdo \: #2}
\newcommand{\resource}[2]{\kresource\: #1 \: \kdo \: #2}
\newcommand{\when}[3]{\kwith\: #1 \: \kwhen\: #2 \: \kdo \: #3}
\newcommand{\alloc}[2]{#1 \coloneqq \kalloc(#2)}
\newcommand{\deref}[1]{[#1]}

\newcommand{\abort}{\mathtt{abort}}
\newcommand{\booleanabort}{\mathbf{abort}}

\DeclareMathOperator*{\bigsprod}{\scalerel*{\circledast}{\sum}}

\newcommand{\GraphofStates}{\mathbf{G(MState)}}
\newcommand{\GraphofSeparatedStates}{\mathbf{G(SState)}}
\newcommand{\SeparationGame}{\mathbf{SGame}}

\newcommand{\PAR}{\textsc{Par}}
\newcommand{\IF}{\textsc{If}}

\newcommand{\CONJ}{\textsc{Conj}}

\newcommand{\WHEN}{\textsc{With}}
\newcommand{\RES}{\textsc{Res}}
\newcommand{\AFF}{\textsc{Aff}}
\newcommand{\SEQ}{\textsc{Seq}}
\newcommand{\FRAME}{\textsc{Frame}}

\newcommand{\STORE}{\textsc{Store}}
\newcommand{\LOAD}{\textsc{Load}}

\newcommand{\fv}{\mathrm{fv}}

\newcommand{\defpred}{\mathrm{def}}

\DeclareMathOperator{\own}{Own}

\newcommand{\emp}{\mathbf{emp}}

\newcommand\eqdef{\stackrel{\mathclap{\normalfont\mbox{\tiny{def}}}}{=}}

\newcommand{\triple}[4]{#1 \vdash\{#2\} \, #3 \, \{#4\}}
\newcommand{\triplestd}{\triple{\G}{P}{C}{Q}}

\newcommand{\parfinmap}{\rightharpoonup_{\mathit{fin}}}

\newcommand{\updmap}[3]{#1[#2 \mapsto #3]}

\newcommand\evalexpr[3]{#1(#2) = #3}

\newcommand{\kArrowLength}{1cm}

\newcommand{\redto}[3]{%
    \begin{tikzcd}[ampersand replacement=\&, column sep=\kArrowLength]%
      {#1} \arrow[r, rightsquigarrow, "#2"] \& {#3}%
    \end{tikzcd}%
  }

\newcommand{\returnto}[3]{%
    \begin{tikzcd}[ampersand replacement=\&, column sep=\kArrowLength]%
      {#1} \arrow[r, rightsquigarrow, tail, "#2"] \& {#3}%
    \end{tikzcd}%
  }

\newcommand{\errorto}[3]{%
    \begin{tikzcd}[ampersand replacement=\&, column sep=\kArrowLength]%
     {#1} \arrow[r, double,  rightsquigarrow, -implies, "#2"] \& {#3}%
    \end{tikzcd}%
  }

\newcommand{\hide}[1]{\mathsf{hide}[#1]}
\newcommand{\inside}[1]{\mathsf{inside}[#1]}
\newcommand{\whentrue}[1]{\mathsf{whentrue}[#1]}
\newcommand{\whenfalse}[1]{\mathsf{whenfalse}[#1]}
\newcommand{\whenabort}[1]{\mathsf{whenabort}[#1]}

\newcommand{\verticalproof}[1]{\begin{array}{c}#1\vspace{-.5em}\\ \vdots\end{array}}

\newcommand{\Board}[1]{\mathsf{Board}_{#1}}
\newcommand{\Polarity}[1]{\mathsf{Pol}_{#1}}
\newcommand{\Plays}[1]{\mathsf{Plays}_{#1}}
\newcommand{\Initial}[1]{\mathsf{Init}_{#1}}

\newcommand{\play}{p} 
\usepackage{authblk}

\newtheorem{thm}{Theorem}

\newtheorem{defin}{Definition}
\newtheorem{definition}[defin]{Definition}
\newtheorem{prop}{Proposition}
\newtheorem{cor}{Corollary}

\title{A Game Semantics of\\ Concurrent Separation Logic}

\author[1]{Paul-Andr\'e Melli\`es}
\author[2]{L\'eo Stefanesco}

\affil[1]{IRIF, CNRS, Universit\'e Paris Diderot}
\affil[2]{\'Ecole Normale Sup\'erieure de  Lyon}

\date{}

\begin{document}
\maketitle

\begin{abstract}
In this paper, we develop a game-theoretic account of concurrent separation logic.
To every execution trace of the Code confronted to the Environment, we associate
a specification game where Eve plays for the Code, and Adam for the Environment.
The purpose of Eve and Adam is to decompose every intermediate machine state
of the execution trace into three pieces: one piece for the Code, one piece
for the Environment, and one piece for the available shared resources.
We establish the soundness of concurrent separation logic by interpreting
every derivation tree of the logic as a winning strategy of this specification game.
\end{abstract}

\section{Introduction}
%
%
%
Concurrent separation logic (CSL) is an extension of Reynold's separation
logic~\cite{Reynolds} formulated by O'Hearn~\cite{OHearn} to establish the
correctness of concurrent imperative programs with shared memory and locks.
This specification logic enables one to establish the good behavior of these
programs in an elegant and modular way, thanks to the frame rule of separation logic.
A sequent of concurrent separation logic
\[
\triple{r_1:P_1,\dots,r_n:P_n}{P}{C}{Q}
\]
consists of a Hoare triple $\{P\}C\{Q\}$ together with a context
$\Gamma=r_1:P_1,\dots,r_n:P_n$ which declares a number of \emph{resource
  variables} $r_k$ (or mutexes) together with the CSL formula~$P_k$ which they
satisfy as invariant.
%
The validity of the program logic relies on a soundness theorem, which states
that the existence of a derivation tree in concurrent separation logic
\begin{center}
\vspace{-2em}
\AxiomC{$\verticalproof{\pi}$}
\UnaryInfC{${\triple{r_1:P_1,\dots,r_n:P_n}{P}{C}{Q}}$}
\DisplayProof
\end{center}
ensures (1) that the concurrent program~$C$ will not produce any race condition
at execution time, and (2) that the program~$C$ will transform every initial
state satisfying $P$ into a state satisfying $Q$ when it terminates, \emph{as
  long as} each resource~$r_k$ allocated in memory satisfies the CSL invariant
$P_k$.
The soundness of the logic was established by Brookes in his seminal
papers on the trace semantics of concurrent separation
logic~\cite{Brookes:a-semantics,Brookes:revisionist}.
His soundness proof was the object of great attention in the community, and it was
revisited in a number of different ways, either semantic \cite{Vafeiadis},
syntactic \cite{mezzo} or axiomatic \cite{views} and formalised in proof
assistants.
One main technical challenge in all these proofs of soundness is to establish
the validity of the concurrent rule:
\begin{center}
\AxiomC{$\triple{\Gamma}{P_1}{C_1}{Q_1}$}
\AxiomC{$\triple{\Gamma}{P_2}{C_2}{Q_2}$}
\RightLabel{\quad Concurrent Rule}
\BinaryInfC{${\triple{\Gamma}{P_1\ast P_2}{C_1\parallel C_2}{Q_1\ast Q_2}}$}
\DisplayProof
\end{center}
and of the frame rule:
\begin{center}
\AxiomC{$\triple{\Gamma}{P}{C}{Q}$}
\RightLabel{\quad Frame Rule}
\UnaryInfC{${\triple{\Gamma}{P\ast R}{C}{Q\ast R}}$}
\DisplayProof
\end{center}
In this paper, we establish the validity of these two rules (and of CSL at
large) based on a new approach inspired by game semantics, which relies on the
observation that the derivation tree $\pi$ of CSL defines a winning strategy
$[\pi]$ in a specification game.
As we will see, the specification game itself is derived from the execution of
the code $C$ and its interaction with the environment (called the frame) using
locks on the shared memory.
The specification game expresses the usual rely-and-guarantee conditions as
winning conditions in an interactive game played between Eve (for the code) and
Adam (for the frame).
In the semantic proofs of soundness, two notions of ``state'' are usually
considered, besides the basic notion \emph{memory state} which describes the
state of the variables and of the heap: (1) the \emph{machine states} which are
used to describe the execution of the code, and in particular include
information about the status of the locks, and (2) the \emph{logical states}
which include permissions and other information invisible at the execution
level, but necessary to specify the states in the logic.
In particular, the tensor product~$\ast$ of separation logic requires
information on the permissions, and 
it is thus defined on logical states, not on
machine states.
The starting point of the paper is the observation that there exists a third
notion of state, which we call \emph{separated state}, implicitly at work in all
the semantic proofs of soundness.
A separated state describes which part of the global (logical) state of the
machine is handled by each component interacting in the course of the execution.
It is defined as a triple $(\state_C, \statevector, \state_F)$ consisting of
\begin{itemize}
\item the logical state $\state_C\in\LogStates$ of the code,
\item the logical state $\state_F\in\LogStates$ of the frame,
\item a function $\statevector:\{r_1,\dots,r_n\}\to \LogStates +\{C,F\}$ which
  tells for every resource variable $r$ whether it is locked and owned by the
  code, $\statevector(r)=C$, locked and owned by the frame, $\statevector(r)=F$,
  or available with logical state $\statevector(r)\in\LogStates$.
\end{itemize}
This leads us to a ``span''
\begin{equation}\label{equation/span}
\begin{tikzcd}[column sep=.7em]
\mbox{machine states} \hspace{.5em}
&&&& \arrow[double,-implies,llll,"{\mathit{refines}}"{swap}]
\hspace{.5em} \mbox{separated states} \hspace{.5em}
\arrow[double,-implies,rrrr,"{\mathit{refines}}"]
&&&&  
\hspace{.5em} \mbox{logical states}
\end{tikzcd}
\end{equation}
where the two notions of machine state and of logical state 
are ``refined'' by the notion of separated state, which conveys 
information about locks (as machine states) and about permissions
(as logical states).
Namely, every separated state $$(\state_C, \statevector,
\state_F)\in\SStates$$ refines the logical state $\mbox{$\bigsprod$}(\state_C, \statevector,\state_F)$
defined by the separation tensor product
\begin{equation}\label{equation/linker}
  \mbox{$\bigsprod$}(\state_C, \statevector,
\state_F)
\quad \eqdef \quad
\state_C * \,\, \Big\{\,\,\bigsprod_{r \in \dom(\statevector)}\statevector(r) \,\, \Big\}\,\, * \state_F
\end{equation}
where $\dom(\statevector)$ denotes the set of resources available in
$\statevector$, in the sense that $\statevector(r)\neq C,F$.
Similarly, every separated state $(\state_C,\statevector,\state_F)$ refines a
machine state $(\memstate, L)$ defined as the memory state $\memstate$
underlying the logical state~(\ref{equation/linker}) just constructed, plus the
set of locked resources $L=\domC(\statevector) \uplus \domF(\statevector)$,
see~\S\ref{sec:separated-states} for details.
In the same way as the notion of logical state is necessary to define the tensor
product $\ast$ of separation logic, and thus to specify the states, the shift
from machine states to separated states is necessary to specify the code, and
the way it interacts with its environment and with its resources.
Our point here is that the formulas $P$ and $Q$ of separation logic in a Hoare
triple $\triplestd$ do not specify the logical state
$\sigma=\mbox{$\bigsprod$}(\state_C, \statevector, \state_F)\in\LogStates$ of
the machine itself, but the \emph{fragment} $\sigma_C$ of this logical
state~$\sigma$ owned by the code~$C$ at the beginning and at the end of the
execution.
The notion of separated state is thus at the very heart of the very concept of
Hoare triple in separation logic.

We follow the following track in the paper.
After discussing the related work, we formulate the two notions of machine
states and of machine instructions
in~\S\ref{sec:machine-states-machine-instructions}.
This enables us to define the notion of execution traces on machine states
in~\S\ref{section/execution-traces} and a number of algebraic operations on
them.
The trace semantics of concurrent programs, and their interpretation as
transition systems, is then formulated in~\S\ref{sec/transition-systems}
and~\S\ref{section/semantics-of-the-language}.
Once the notion of machine state has been used to describe the trace semantics
of the language, we move to the logical side of the span, and formulate the
notions of logical state in \S\ref{sec:logical-states} and the notion of
separated state in~\S\ref{sec:separated-states}.
In~\S\ref{sec:specification-games}, we explain how to associate to every
execution trace $t$ a specification game played on the paths of the graph of
separated states, which is defined in~\S\ref{sec:graphs}. The moves of those
games express the ownership discipline enforced by separation logic, and in
particular the discipline associated to the locks in concurrent separation
logic.
Finally, we show in~\S\ref{sec:soundness} that CSL is sound by proving that
every derivation tree of the logic defines a strategy, which lifts each step of
the Code of an execution trace into the graph of separated states.
%



\section{Related Work}
Several proofs of soundness have already been given for concurrent separation
logic.
%
%
The first proof of correctness was designed by Brookes
in~\cite{Brookes:a-semantics,Brookes:revisionist} using semantic ideas.
In his proof, every program~$C$ is interpreted as a set of ``action traces'',
defined as finite or infinite sequences of ``actions'' that look like:
\begin{center}
$\mathtt{read~} 71~\mathtt{from~} x$,\; $\mathtt{read~} 36~\mathtt{from}~y$,\; $\mathtt{acquire~lock~r}$, \dots. 
\end{center} 
An interesting feature of the model is that these action traces do not mention
(at least explicitly) the machine states produced by the Code at execution time.
The environment is taken into account through the existence of \emph{non
  sequentially consistent traces} such as
\begin{center}
$\mathtt{write~} 89~\mathtt{in}~x$, $\mathtt{read~} 14~\mathtt{from}~x$
\end{center}
in the model.
The idea is that the Environment presumably changed the value of the variable
$x$ between the two actions of the Code.
Separation in the logic enables one to decompose actions traces into \emph{local
  computations}, in order to reflect the program's subjective view of the
execution.
%


Vafeiadis gave another proof of correctness~\cite{Vafeiadis} based on more
directly operational intuitions.
In his proof, the Code is interpreted as a \emph{transition system} whose
vertices are pairs $(C,\state)$ consisting of the Code $C$ and of the state
$\state$ of the memory, and where edges are execution steps.
The core of the soundness proof is that each step of the execution preserves a
decomposition of the heap into three parts, which correspond respectively to the
Code, the resources, and the Frame.
The proof is done by induction on the derivation tree $\pi$ establishing the
triple $\triplestd$ in concurrent separation logic.
The idea of using separated states thus comes from Vafeiadis' proof, which is
the closest to ours.
One difference, however, besides the game-theoretic point of view we develop, is
that we have a more \emph{intensional} description of separated states, provided
by the function $\statevector$ which tracks the states of each of the available
locks.

%
%

In contrast to the semantic proofs mentioned above, Balabonski, Pottier and
Protzenko \cite{mezzo} developed a purely syntactic proof of correctness for
Mezzo, a functional language equipped with a type-and-capability system based on
concurrent separation logic. The soundness of the logic follows in their
approach from a \emph{progress} and a \emph{preservation} theorem on the type
system of Mezzo.

Our focus in this work is to develop a game-theoretic approach to concurrent
separation logic. For that reason, we prefer to keep the logic as well as the
concurrent language fairly simple and concrete. In particular, we do not
consider more recent, sophisticated and axiomatic versions of the logic, like
Iris~\cite{iris1,iris3}.

\section{Machine states and machine instructions}\label{sec:machine-states-machine-instructions}
The purpose of this section is to introduce the notions of \emph{machine state}
and of \emph{machine instruction} which will be used all along the paper.
%
We suppose given countable sets $\Var$ of variable names, $\Val$ of values,
$\Loc \subseteq \Val$ of memory locations, and $\RVar$ of resources.
In practice, $\Loc = \mathbb{N}$ and $\Val = \mathbb{Z}$.

\begin{defin}[Memory state]
  A memory state $\memstate$ is a pair $(s,h)$ of partial functions with finite
  domains $s: \Var \parfinmap \Val$ and $h: \Loc \parfinmap \Val$ called the
  stack~$s$ and the heap~$h$ of the memory state~$\memstate$.
  The set of memory states is denoted $\State$.
  The domains of the partial function $s$ and of $h$ are noted
  $\vdom(\memstate)$ and $\hdom(\memstate)$ respectively, and we write
  $\dom(\memstate)$ for their disjoint union.
\end{defin}

\begin{defin}[Machine state]
  A machine state $\mstate=(\memstate,L)$ is a pair consisting of a memory state
  $\memstate$ and of a subset of resources $L\subseteq\RVar$, called the lock
  state, which describes the subset of locked resources in $\mstate$. The set of
  machine states is denoted $\MStates$.
\end{defin}
A \emph{machine step} is defined as a labelled transition between machine
states, which can be of two different kinds:
\[
  \returnto{\mstate}{m}{\mstate'}
  \quad\quad\quad\quad\quad
  \errorto{\mstate}{m}{\mstate'}
\]
depending on whether the instruction $m\in\Instr$ has been executed successfully
(on the left) or it has produced a runtime error (on the right).
We write $m:\redto{\mstate}{}{\mstate'}$ when we do not want to specify whether
the instruction has produced a runtime error.
The \emph{machine instructions} which label the machine steps are defined below:
\begin{align*}
  m &::=   x \coloneqq E
      \mid x \coloneqq \deref{E}
      \mid \deref{E} \coloneqq E'
      \mid \knop
      \mid x \coloneqq \kalloc(E)
      \mid \kdispose(E)
      \mid P(r)
      \mid V(r)
\end{align*}
where $x\in\Var$ is a variable, $r\in\RVar$ is a resource variable,
and $E,E'$ are arithmetic expressions with variables.
Typically, the instruction $x\coloneqq E$ assigns to the variable~$x$
the value $E(\mu)$ of the expression $E$ in the memory state~$\mu$,
the instruction $P(r)$ locks the resource variable $r$ when it is available,
while the instruction $V(r)$ releases it when it is locked,
as described below:
$$
\begin{small}
 \prftree{\evalexpr{E}{\memstate}{v}}
  {\returnto{(\memstate, L)}{x \coloneqq E}{(\updmap{\memstate}{x}{v}, L)}}
  \quad
    \prftree{r\notin{}L}
  {\returnto{(\memstate, L)}{P(r)}{(\memstate, L\uplus\{r\})}}
   \quad
    \prftree{r\notin{}L}
  {\returnto{(\memstate, L\uplus\{r\})}{V(r)}{(\memstate, L)}}
  \end{small}
$$
Thanks to the inclusion $\Loc \subseteq \Val$, an expression $E$
may also denote a location.
In that case, $[E]$ refers to the value of the location $E$ in memory.
The instruction $\knop$ (for no-operation) does not alter the logical state,
while $x \coloneqq \kalloc(E)$ allocates (in a non-deterministic way) some
memory space on the heap, initializes it with the value of the expression $E$,
and returns the address of the location to the variable~$x$, while
$\kdispose(E)$ deallocates the location with address $E$.

It will be convenient in the sequel to write $\lock^+(m)$ for the set of locks
which are taken by an instruction $m$, that is, $\lock^+(m)=\{r\}$ if $m=P(r)$
and $\lock^+(m)=\emptyset$ otherwise; $\lock^-(m)$ is the set of locks which are
released by the instruction $m$, that is, $\lock^-(m)=\{r\}$ if $m=V(r)$ and
$\lock^-(m)=\emptyset$ otherwise.

\section{Execution traces}\label{section/execution-traces}
Now that the notion of machine state has been introduced, the next step towards
the interpretation of programs is to define the notion of execution trace, with
two kinds of transitions: the even transitions ``played'' by the Code, and the
odd transitions ``played'' by the Environment.

\begin{defin}[Traces]\label{def/traces}
  A \emph{trace} $t$ is a sequence of machine states
  \[
    \mstate_1\xrightarrow{env} \mstate_2 \xrightarrow{m_1} 
    \mstate_3\xrightarrow{env}
    \ldots\xrightarrow{env}
    \mstate_{2p}\xrightarrow{m_p}\mstate_{2p+1}\xrightarrow{env}\mstate_{2p+2}
  \]
  whose even transitions
\[
  \mstate_{2k}\, \xrightarrow{\;\; m_k \;\;} \, \mstate_{2k+1}
  \quad\quad\quad \mbox{$1\leq k\leq p$}
  \]
  are labelled by an instruction $m_k\in\Instr$ such that $ \redto
  {\mstate_{2k}}{m_k}{\mstate_{2k+1}}$ and whose last transition is played by
  the environment. The set of traces is denoted by $\Traces$.
\end{defin}
\noindent
%
%
We write $\source{t}=\mstate_1$ and $\target{t}=\mstate_{2p+2}$ for the initial
and the final states of a trace $t\in\Traces$, respectively.
%
%
The length $\length{t}=p$ is defined as the number of Code transitions in the
trace, and
\[
  \step{t}{k} \quad = \quad \mstate_{2k}\, \xrightarrow{\;\; m_k \;\;} \, \mstate_{2k+1}
\]  
denotes the $k$-th even transition of the trace~$t$, for $1\leq
k\leq\length{t}$.
Observe that a trace~$t$ always starts and stops by an Environment transition,
and that its number of transitions is equal to $2\times\length{t}+1$.
We point out the following fact which we will often use in our proofs and
constructions:
\begin{prop}
  A trace $t\in\Traces$ is characterized by its initial state~$\source{t}$ and
  by its final state~$\target{t}$, together with the sequence of Code
  transitions $\step{t}{k}$ for $1\leq k\leq \length{t}$.
\end{prop}

We introduce now a number of important algebraic constructions on execution
traces, whose purpose is to reflect at the level of traces the sequential and
parallel composition of programs.

\begin{defin}[Sequential composition]
  Given two traces $t_1, t_2\in\Traces$ such that
  $\partial_1(t_1)=\partial_0(t_2)$, one defines $t_1\cdot t_2\in\Traces$ as the
  trace of length $\length{t_1}+\length{t_2}$ with initial state
  $\partial_0(t_1)$ and final state $\partial_1(t_2)$, and with even transitions
  defined as 
  \[
    \step{(t_1 \cdot t_2)}{k} =
    \begin{cases*}
      \step{t_1}{k} & if $1 \leq k \leq p$,\\
      \step{t_2}{k-p} & if $p+1 \leq k \leq p+q$.
    \end{cases*}
  \]
\end{defin}
\begin{defin}[Restriction]\label{def/restriction}
  Let $\Traces_p$ denote the set of traces of length $p$. Every increasing
  function $f:\{1,...,p\}\to \{1,...,q\}$ induces a restriction function
  \[
  \pullback{f} : \Traces_{\,q} \longrightarrow \Traces_{\,p}
  \]
  which transports a trace $t$ of length $q$ to a coinitial and cofinal trace
  $\pullback{f}(t)$ of length $p$
  \[
  \source{\pullback{f}(t)}=\source{t}\quad\quad\quad \target{\pullback{f}(t)}=\target{t}
  \]
  defined by the instructions $\step{\pullback{f}(t)}{k} = \step{t}{f(k)}$ for
  $1\leq k\leq p$.
\end{defin}
  
  \begin{defin}[Shuffle]\label{def/shuffle}
A shuffle of two natural numbers $p\in\Nat$ and $q\in\Nat$ is a monotone bijection
  $
    \omega: \{1,\ldots ,p\} \uplus \{1, \ldots, q\} \to \{1, \ldots, p+q \}.
    $
The set of shuffles of $p$ and $q$ is denoted $\Shuffles{p}{q}$.
  \end{defin}
\noindent
Every shuffle $\omega\in\Shuffles{p}{q}$ induces a pair of increasing functions
  \[
  \omega_1 : \{1,...,p\} \to \{1, \ldots, p+q \}
  \quad \text{ and } \quad
  \omega_2  :  \{1,...,q\}  \to  \{1, \ldots, p+q \}
  \]
defined by restricting $\omega$ to $\{1,...,p\}$ and to $\{1,...,q\}$,
respectively.
From this follows immediately that
\begin{prop}
Every shuffle $\omega\in\Shuffles{p}{q}$ induces a function
  \[
  \pullback{\omega} : \Traces_{p+q} \longrightarrow \Traces_p\times\Traces_q
  \]
which transports a trace $t$ of length $p+q$ to the pair 
$(\pullback{\omega_1}(t),\pullback{\omega_2}(t))\in\Traces_p\times\Traces_q.$
\end{prop}

\begin{defin}
  The parallel composition $t_1\parallel t_2$ is the set of traces $t\in\Traces$
  such that $\pullback{\omega}(t)=(t_1,t_2)$ for some shuffle
  $\omega\in\Shuffles{\length{t_1}}{\length{t_2}}$.
\end{defin}
\noindent
Note that every trace $t$ in $t_1\parallel t_2$ satisfies
$\length{t}=\length{t_1}+\length{t_2}$ and more importantly, that the parallel
composition $t_1\parallel t_2$ of two traces $t_1$ and $t_2$ is empty whenever
the two traces $t_1$ and $t_2$ are not coinitial and cofinal.
The purpose of our last construction $\hide{r}$ is to ``hide'' the name of a
resource variable $r\in\RVar$ in an execution trace.
\begin{defin}
  The function $\hide{r} : \Traces\to\Traces$ transforms every trace by applying
  the function
  \[
    (\memstate, L) \longmapsto (\memstate, L\setminus\{r\}) \quad : \quad
    \MStates \longrightarrow \MStates
  \]
  to each machine state of the original trace, and the function
    \[
    \begin{array}{ccl}
      m & \longmapsto & \begin{cases*}
        \,\, \knop & if $m = P(r)$ or $V(r)$\\
        \,\, m     & otherwise
      \end{cases*}
    \end{array}
    \quad : \quad  \Instr \longrightarrow \Instr
  \]
  to the instructions of the trace.
\end{defin}

\section{Transition Systems}\label{sec/transition-systems}
At this stage, we are ready to introduce the notion of transition system which
we will use in order to describe the traces generated by a program of our
concurrent language.
Among these execution traces, one wishes to distinguish (1) the traces which
terminate and return from (2) the other traces which are not yet finished or
terminate and abort.
This leads us to the following definition of transition system:

\begin{defin}[Transition Systems]
  A \emph{transition system} $\mathbf{T}=(T,\returntraces{T})$ is a set of
  traces $T \subseteq \Traces$ closed under prefix, together with a subset
  $\returntraces{T}\subseteq T$, whose traces are said to \emph{return}.
  \end{defin}
\noindent
We explain below how to lift to transition systems the algebraic operations
defined on traces in the previous section~\S\ref{section/execution-traces}.

\begin{defin}\label{def:seq-comp}
  The \emph{sequential composition} of two transition systems $\mathbf{T}$ and
  $\mathbf{T}'$, is defined as the transition system $\mathbf{T} \sequential
  \mathbf{T}'$ below:
   \begin{align*}
     T \sequential T' \;\;&=\;\;  T \, \cup \, \{t\cdot t' \mid t \in \returntraces{T}, t' \in T' \text{ and } \target{t} = \source{t'}\}
     \\
     \returntraces{T \sequential T'} \;\;&=\;\; \{t \cdot t'\mid t \in \returntraces{T}, t' \in \abs{T'} \text{ and } \target{t} = \source{t'}\}
   \end{align*}
\end{defin}

\begin{defin}\label{def:par-comp-trace}
  The \emph{parallel composition} of two transition systems $\mathbf{T}$ and
  $\mathbf{T}'$, is defined as the transition system $\mathbf{T} \parallel
  \mathbf{T}'$ below:
      \[
    {T_1 \parallel T_2} \;\;=\;\;  \bigcup_{t_i\in T_i}{t_1 \parallel t_2}
    \quad\quad\quad\quad
    \returntraces{T_1 \parallel T_2} \;\; = \;\; \bigcup_{t_i \in \returntraces{T_i}}{t_1 \parallel t_2}
  \]
\end{defin}

\begin{defin}
  The transition system~$\hide{r}(\mathbf{T})$ associated to a transition
  system~$\mathbf{T}$ and to a lock~$r\in\RVar$ is defined as follows:
  \[
    \hide{r}(T) \; = \; \{\hide{r}(t) \mid t \in T \}
    \quad\quad\quad
    \returntraces{\hide{r}(T)} \; = \; \{\hide{r}(t) \mid t \in \returntraces{T} \}.
  \]
\end{defin}
Note that every instruction $m\in\Instr$ induces a transition system $\sem{m}$
defined in the following way:
\[ \begin{array}{ccc}
  \sem{m} & = & 
  \{ \mstate_1 \xrightarrow{env} \mstate_2 \xrightarrow{m} \mstate_3 \xrightarrow{env} \mstate_4 \mid \redto{\mstate_2}{m}{\mstate_3}\}
  \\
\returntraces{\sem{m}} & = &
 \{ \mstate_1 \xrightarrow{env} \mstate_2 \xrightarrow{m} \mstate_3 \xrightarrow{env} \mstate_4 \mid \returnto{\mstate_2}{m}{\mstate_3}\}
\end{array}
\]
The intuition is that the program interpreted by $\sem{m}$ executes the
instruction $m$ after the environment has made the transition $\mstate_1
\xrightarrow{env} \mstate_2$ and returns when the machine step $\mstate_2
\xrightarrow{m} \mstate_3$ is succesful, and does not abort.
The following algebraic operation on transition systems reflects the
computational situation of a program taking a lock $r$ before executing, and
releasing the lock $r$ in case the program returns.


\begin{defin}
  The transition system~$\inside{r}(\mathbf{T})$ associated to a transition
  system~$\mathbf{T}$ and to a lock $r\in\RVar$ is defined as follows:
  \[
    \inside{r}(\mathbf{T}) = \sem{P(r)};\mathbf{T};\sem{V(r)}.
  \]
\end{defin}
The following operation on transition systems will enable us to interpret
conditional branching on concurrent programs.
\begin{defin}
  The transition system $\whentrue{P}(\mathbf{T})$ associated to a transition
  system~$\mathbf{T}=(T,\returntraces{T})$ and a predicate
  $P:\MStates\to\{\True,\False,\booleanabort\}$ on memory states is defined as
  follows:
\[
\begin{array}{ccc}
\whentrue{P}(\mathbf{T})  & = &  \{ t\in T \, | \, P(\sourcecode{t}) = \True \}
\\
\returntraces{\whentrue{P}(\mathbf{T})} & = & \{ t\in \returntraces{T} \, | \, P(\sourcecode{t}) = \True \}
\end{array}
\]
where $\sourcecode{t}=\mstate_2$ denotes the first state played by Code in the
trace $t$.
\end{defin}
The transition system $\whenfalse{P}(\mathbf{T})$ is defined similarly, by
replacing $\True$ by $\False$ in the definition.
A subtle but important aspect of the interpretation of conditional branching in
the language is that the evaluation of a boolean expression~$B$ may not succeed,
typically because one of its variables $x\in\Var$ is not allocated.
In that case, the evaluation produces an exception which is then handled by the
operating system.
This $\abort$ case is handled in our trace semantics by the definition of a
dedicated transition system called $\whenabort{P,C}$, whose construction is
detailed in the Appendix\cite{Appendix}.

\section{Trace semantics of the concurrent language}\label{section/semantics-of-the-language}
Now that we have defined the basic operations on transition systems, 
we are ready to define the operational and interactive semantics of
our concurrent language.
%
The language is constructed with
Boolean expressions~$B$, arithmetic
expressions~$E$ and commands~$C$,
using the grammar below:
\begin{align*}
  B &::=   \True
      \mid \False
      \mid B \wedge B'
      \mid B \vee B'
      \mid E = E' \\
  E &::=   0
      \mid 1
      \mid \ldots
      \mid x
      \mid E + E'
      \mid E * E'\\
  C &::=   x \coloneqq E
      \mid x \coloneqq \deref{E}
      \mid \deref{E} \coloneqq E'
      \mid C ; C'
      \mid C_1 \parallel C_2
      \mid \kskip\\
    &\mid \while{B}{C}
      \mid \resource{r}{C}
      \mid \when{r}{B}{C}\\
    &\mid \ifte{B}{C_1}{C_2}
      \mid x \coloneqq \kalloc(E)
      \mid \kdispose(E)
\end{align*}
%
The parallel composition operator~$C_1\parallel C_2$ enables the two programs $C_1$ and $C_2$ to interact concurrently through \emph{mutexes} called \emph{resources}.
A resource~$r$ is declared using $\kresource~r$ and acquired using $\when{r}{B}{C}$, which waits for the Boolean expression $B$ to be true in order to proceed. Of course, a mutex can be held by at most one execution thread at any one time.
In the semantic approach we are following,
every command $C$ is translated into a transition system $\sem{C}$ which
describes the possible interactive executions of $C$, and whether they return.
\begin{center}
\begin{tikzcd}[column sep=3em]
\mbox{Code } C\quad
\ar[rr,"\text{translation}"]
& {} &
\quad\mbox{Transition system } \sem{C}
\end{tikzcd}
\end{center}
The interpretation $\sem{C}$ is defined by structural induction on the syntax of
the command~$C$.
To each leaf node $C$, one associates an instruction~$m\in\Instr$
\[
x \coloneqq E
\mid x \coloneqq \deref{E}
\mid \deref{E} \coloneqq \deref{E'}
\mid \knop
\mid \alloc{x}{E}
\mid \kdispose(E)
\]
which defines the transition system
$
  \sem{C} \eqdef \sem{m}.
  $
The semantics of non-leaf commands is then defined
using the algebraic operations on transition systems
introduced in~\S\ref{sec/transition-systems}:
\[
\sem{C \parallel C'} \, \eqdef \, \sem{C} \parallel \sem{C'},
\quad\quad
\sem{C ; C'} \, \eqdef \, \sem{C} \sequential \sem{C'},
\]
\[
\sem{\kresource\,{r}\,\kdo\,{C}} \, \eqdef \, \hide{r}\,\big(\sem{C}\big),\]
\[
  \sem{\when{r}{B}{C}} \,\,\eqdef\,\,
  \whentrue{B}\Big( \inside{r}\big(\sem{C}\big) \Big) \,\, \cup \,\,
  \whenabort{B,C'}
\]
where $C'= \when{r}{B}{C}$ in the last part of the definition, and finally
\begin{align*}
\sem{\ifte{B}{C_1}{C_2}} \;\; &\eqdef \;\;
 \whentrue{B}\big(\sem{\knop}\big) \sequential \sem{C_1}
\;\cup\; \whenfalse{B}\big(\sem{\knop}\big) \sequential \sem{C_2}
\\
&\quad \cup \;\whenabort{B,\ifte{B}{C_1}{C_2}},
\end{align*}
and the while loop
\[
\sem{\while{B}{C}} \; \eqdef \; \bigcup_{n \geq 0} \,\, F^n(\emptyset)
\]
is defined as the least fixpoint 
of the continuous function $F : \Trans \to  \Trans$ 
below:
\begin{align*}
F(\mathbf{T})
\quad=\quad
&\whentrue{B}\big(\sem{\knop}\big) \sequential \sem{C}\sequential\mathbf{T}
\;\; \cup \;\;  \whenfalse{B}\big(\sem{\knop}\big) \;\; \cup
\\
&\whenabort{B,\while{B}{C}}.
\end{align*} 

\section{Logical States}\label{sec:logical-states}
As we explained in the introduction, reasoning about concurrent programs in
separation logic requires introducing an appropriate notion of \emph{logical
  state}, including information about permissions.
The version of concurrent separation logic we consider is almost the same as in
its original formulation by O'Hearn and
Brookes~\cite{OHearn,Brookes:a-semantics}.
One difference is that we benefit from the work
in~\cite{Bornat:permissions,Bornat:sep,Bornat:hoare} and use the permissions and
the $\own_p(x)$ predicate in order to handle the heap as well as variables in the stack.
So, we suppose given an arbitrary partial cancellative commutative monoid
$\Perm$ that we call the \emph{permission monoid}, following
\cite{Bornat:permissions}.
We require that the permission monoid contains a distinguished element~$\top$
which does not admit any multiple, ie. $ \forall x\in\Perm, \top \cdot x \text{
  is not defined.} $ The idea is that the permission~$\top$ is required for a
program to write somewhere in memory.
The property above ensures that a piece of state cannot be written and accessed
(with a read or a write) at the same time by two concurrent programs, and
therefore, that there is memory safety and no data race in the semantics.
The set $\LStates$ of logical states is defined in a similar way as the set
$\State$ of memory states, with the addition of permissions:
\[
  \LStates \;=\; (\Var \parfinmap\Val \times \Perm) \times
  (\Loc \parfinmap\Val \times \Perm)
\]
%
One main benefit of permissions is that they enable us to define a
\emph{separation tensor product} $\state * \state'$ between two logical states
$\state$ and $\state'$.
When it is defined, the logical state $\state * \state'$ is defined as a partial
function with domain
$$\dom(\state * \state) = \dom(\state) \cup \dom(\state')$$
in the following way, for $a\in\Var\amalg\Loc$:
\[
  \state*\state'(a) =
  \begin{cases*}
    \state(a) & if $a \in \dom(\state)\setminus\dom(\state')$\\
    \state'(a) & if $a \in \dom(\state')\setminus\dom(\state)$\\
    (v,p\cdot{}p') & if $\state(a) = (v, p)$ and $\state'(a) = (v,p')$\\
  \end{cases*}
\]
The tensor product $\state * \state'$ of the two logical states $\state$ and
$\state'$ is not defined otherwise.
In other words, if the tensor product is well defined, then the memory states
underlying $\state$ and $\state'$ agree on the values of the shared variables
and heap locations.
The syntax and the semantics of the formulas of Concurrent Separation Logic is
the same as in Separation Logic.
%
%
The grammar of formulas is the following one:
\begin{align*}
  P,Q,R,J \Coloneqq \; &\emp \mid \True \mid \False \mid P \vee Q \mid
                         P \wedge Q \mid \neg P
  \mid\; \forall X. P
         \mid \; \exists X. P\\
  \mid\; &P * Q \mid \own_p(x) \mid E_1 \mapsto^{p} E_2
\end{align*}
The semantics of the formulas is expressed as the satisfaction predicate $\state
\vDash P$ defined in Figure~\ref{fig:CSL-predicate-semantics}.
\begin{figure}[!t]
\begin{align*}
  \s \vDash \own_p(x) \;&\Longleftrightarrow\; \exists v\in\Val, \s(x) = (v, p)\\
  \s \vDash E_1 = E_2 \;&\Longleftrightarrow\; \sem{E_1} = \sem{E_2} \wedge \fv(E_1=E_2) \subseteq \vdom(h)\\
  \s \vDash P \Rightarrow Q \;&\Longleftrightarrow\; ( \s \vDash P ) \Rightarrow ( \s \vDash Q )\\
  \s \vDash P \wedge Q \;&\Longleftrightarrow\; \s \vDash P \text{ et } \s \vDash Q\\
  \s \vDash P*Q \;&\Longleftrightarrow\; \exists \s_1 \s_2, \, \s = \s_1 * \s_2 \text{ et } \s_1 \vDash P \text{ et } \s_2 \vDash Q
\end{align*}
\vspace{-1.6em}
\caption{Semantics of the predicates of concurrent separation logic}
\label{fig:CSL-predicate-semantics}
\end{figure}
The proof system underlying concurrent separation logic is a sequent calculus on
sequents defined as Hoare triples of the form
\[
  \triplestd,
\]
where $C\in\Code$, $P$, $Q$ are predicates, and $\Gamma$ is a context, defined
as a partial function with finite domain from the set~$\RVar$ of resource
variables to predicates.
Intuitively, the context $\Gamma=r_1:J_1,\dots,r_k:J_k$ describes the invariant
$J_i$ satisfied by the resource variable $r_i$.
The purpose of these resources is to provide the fragments of memory shared
between the various threads during the execution.
The inference rules are given in Figure~\ref{fig:inference-rules}.
The inference rule $\RES$ associated to $\resource{r}{C}$ moves a piece of
memory which is owned by the Code into the shared context~$\Gamma$, which means
it can be be accessed concurrently inside $C$.
However, the access to said piece of memory is mediated by the $\kwith$
construct, which grants temporary access under the condition that one must give
it back (rule $\WHEN$).
Notice that in the rule \CONJ, the context $\Gamma=r_1:J_1,\dots,r_k:J_k$ is
required to be \emph{precise}, in the sense that each of the predicates $J_i$ is
precise.
\begin{defin}[Precise predicate]
  A predicate $P$ is \emph{precise} when, for any
  $\state\in\LStates$, there exists at most one $\state'\in\LStates$ such that
  $\exists \state'',\; \state=\state'*\state''$ and $\state' \vDash P$.
\end{defin}

\begin{figure*}[!h] 
  \centering
  \small
\begin{mathpar}
  \prfbyaxiom{\AFF}{\triple{\Gamma}{\own_\top(x) * X = E}{x \coloneqq
      E}{\own_\top(x) * x = X}}
  \and
\prfbyaxiom{\STORE}
{\triple
  {\Gamma}
  {E \mapsto -}
  {\deref{E} \coloneqq E'}
  {E \mapsto E'}}
  \and
\prftree[r]{\LOAD}
{x \notin \fv(E)}
{\triple
  {\Gamma}
  {E \mapsto^{p} v * \own_{\top}(x)} 
  {x \coloneqq \deref{E}}
  {E \mapsto^{p} v * \own_{\top}(x) * x = v}}
  \and
  \prftree[r]{\SEQ}{\triplestd}{\triple{\Gamma}{Q}{C'}{R}}{\triple{\Gamma}{P}{C;C'}{R}}
\and
\prftree[r]{\IF}
     {P \Rightarrow \defpred(B)}
     {\triple{\Gamma}{P\wedge B}{C_1}{Q}} 
     {\triple{\Gamma}{P\wedge \neg B}{C_2}{Q}} 
     {\triple{\Gamma}{P}{\ifte{B}{C_1}{C_2}}{Q}}
     \and
     \prftree[r]{\CONJ}{\Gamma\text{ is precise}}{\triple{\Gamma}{P_1}{C}{Q_1}}{\triple{\Gamma}{P_2}{C}{Q_2}}{\triple{\Gamma}{P_1
      \wedge P_2}{C}{Q_1 \wedge Q_2}}
  \and
  \prftree[r]{\RES}{\triple{\Gamma,
      r:J}{P}{C}{Q}}{\triple{\Gamma}{P*J}{\resource{r}{C}}{Q*J}}
  \and
  \prftree[r]{\WHEN}{P\Rightarrow \defpred(B)}{\triple{\Gamma}{(P*J)\wedge
      B}{C}{Q*J}}{\triple{\Gamma, r:J}{P}{\when{r}{B}{C}}{Q}}
  \and
  \prftree[r]{\PAR}{\triple{\Gamma}{P_1}{C_1}{Q_1}}{\triple{\Gamma}{P_2}{C_2}{Q_2}}{\triple{\Gamma}{P_1*P_2}{C_1\parallel
      C_2}{Q_1 * Q_2}}
  \and
  \prftree[r]{\FRAME}
  {\triple{\Gamma}{P}{C}{Q}}
  {\triple{\Gamma}{P*R}{C}{Q*R}}
\end{mathpar}
\normalsize
\vspace{-1.6em}
\caption{Inference rules of Concurrent Separation Logic}
\label{fig:inference-rules}
\end{figure*}

\section{Separated states}\label{sec:separated-states}
We now introduce our third notion of state, which display which region of
(logical) memory belongs to the Code, which region belongs to the Frame, and
which region is shared.
We suppose given a finite set~$\RVar$ of resource variables.
\begin{defin}\label{def:separated-states}
  The \emph{separated states} are the triples
    \[
    (\state_C, \statevector, \state_F) \in \LStates \times (\RVar \to \LStates+\{C,F\}) \times \LStates
  \]
such that the state below is defined:
\begin{equation*}
  \state_C * \,\, \Big\{\,\,\bigsprod_{r \in \dom(\statevector)}\statevector(r) \,\, \Big\}\,\, * \state_F
\end{equation*}
where 
\vspace{-2em}
\begin{align*}
    \dom(\statevector)  &= \{r\in\RVar \mid
                              \statevector(r)\in\LStates\hspace{0.08em}\},
    \\
    \domC(\statevector) &= \{r\in\RVar \mid \statevector(r)=C\},
    \\
    \domF(\statevector) &= \{r\in\RVar \mid \statevector(r)=F\}.
\end{align*}
\end{defin}
\noindent
We say that a separated state $(\state_C, \statevector, \state_F)$ combines into
a machine state $\mstate = (\memstate, L)$ precisely when both $L =
\domC(\statevector)\uplus\domF(\statevector)$ and the memory
state~$\memstate\in\State$ is equal to the image of
\begin{equation}\label{eqn:sep-state-product}
  \state_C * \,\, \Big\{\,\,\bigsprod_{r \in \dom(\statevector)}\statevector(r) \,\, \Big\}\,\, * \state_F
  \quad \in \quad \LStates
\end{equation}
under the function $U:\LStates\to\State$ which forgets the permissions.
Note that by definition, every separated state $(\state_C, \statevector,
\state_F)$ combines into a unique machine state, which we write for concision $
(\memstate,L) = \mbox{$\bigsprod$}(\state_C, \statevector, \state_F). $

\section{The graphs of machine and separated states}\label{sec:graphs}
In this section, we introduce the two labeled graphs
$\GraphofStates$ and $\GraphofSeparatedStates$
of machine states and of separated states,
and construct a graph homomorphism
\begin{equation}\label{equation/Sgraph-to-Mgraph}
\mbox{$\bigsprod$} \quad : \quad \GraphofSeparatedStates \longrightarrow \GraphofStates
\end{equation}
which maps every separated state $(\state_C, \statevector, \state_F)$ to its
combined machine state $(\sigma,L)$, in the way described in the introduction.

\begin{defin}
  The \emph{graph of machine states} $\GraphofStates$ is the graph whose
  vertices are the machine states $\mstate\in\MStates$ and whose edges are
  either Code or Environment transitions of the following kind:
  \begin{itemize}
  \item a Code transition $\mstate \xrightarrow{m} \mstate'$ for every machine
    step $\redto{\mstate}{m}{\mstate'}$,
  \item an Environment transition $\mstate \xrightarrow{env} \mstate'$ for every
    pair $\mstate,\mstate' \in \MStates$ of machine states, and where $env$ is
    just a tag indicating that the transition has been fired by the Environment.
  \end{itemize}
\end{defin}
Note that a trace $t\in\Traces$ (see Def.~\ref{def/traces}) is the same thing as
an alternating path starting and ending with an Environment edge in the graph
$\GraphofStates$.

\begin{defin}\label{def:separation-graph}
  The \emph{graph of separated states} $\GraphofSeparatedStates$ is the graph
  whose vertices are the separated states and whose edges are either Eve moves
  or Adam moves of the following kind:
  \begin{itemize}
  \item Eve moves of the form
    \[
      (\state_C, \statevector, \state_F) \xrightarrow{\,\,m\,\,} (\state'_C, \statevector', \state_F)
    \]
    labeled by an instruction $m\in\Instr$ such that
    \[
      \returnto{\bigsprod(\state_C, \statevector, \state_F)}{m}{\bigsprod(\state'_C, \statevector', \state_F)}
    \]
    between machine states, and such that the following conditions on locked
    resources are moreover satisfied:
    \begin{align*}
      \forall r\notin\lock(m),\; &\statevector(r) = \statevector'(r),\\
      \forall r\in\lock^+(m),\; &r \in \dom(\statevector) \wedge r\in\domC(\statevector'),\\
      \forall r\in\lock^-(m),\; &r\in\domC(\statevector) \wedge r\in\dom(\statevector');
    \end{align*}
  \item Adam moves of the form
    \[
      (\state_C, \statevector, \state_F) \xrightarrow{\,\,env\,\,}
      (\state_C, \statevector', \state'_F)
    \]
    where $env$ is just a tag, and moreover
    \[
      \domC(\statevector') = \domC(\statevector).
    \]
  \end{itemize}
\end{defin}
\noindent
The definition of the vertices and of the edges of the graph of separated states
$\GraphofSeparatedStates$ is designed to ensure that there exists a graph
homomorphism~(\ref{equation/Sgraph-to-Mgraph}) which maps every Eve move to a
Code transition, and every Adam move to an Environment transition.
The graph homomorphism~(\ref{equation/Sgraph-to-Mgraph}) enables us to study how
an execution trace~$t\in\Traces$ defined as a path in~$\GraphofStates$ may be
``refined'' into a \emph{separated execution trace}~$\play$ living in the graph
of $\GraphofSeparatedStates$ of separated states, and such that $t=\bigsprod
\play$.
In that situation, we use the following terminology:
\begin{definition}
  We say that a path~$\play$ in the labeled graph $\GraphofSeparatedStates$
  \emph{combines} into a trace~$t\in\Traces$ in the labeled graph
  $\GraphofStates$ when $t=\bigsprod \play$.
\end{definition}
\noindent
Note that a path $\play$ which combines into a trace $t\in\Traces$ is alternated
between Eve and Adam moves, and that it starts and stops with an Adam move.

\section{Separation games}\label{sec:specification-games}
In this section, we explain how to associate to every trace $t\in\Traces$ a
\emph{separation game} $\SeparationGame(t)$ on which Eve and Adam interact and
try to ``justify'' every transition played in the execution trace $t$ by the
Code or by the Environment, by lifting it to a separated execution trace $\play$
which combines into $t$.
\begin{defin}[Game]
  A \emph{game}~$A$ is a triple $A=(\Board{A}, \Polarity{A}, \Plays{A})$
  consisting of a graph $\Board{A}=(V,E,\source,\target)$ with source and target
  functions $\source,\target : E \to V$, and whose edges are called moves: of a
  function $ \Polarity{A} : E \to \{-1,+1\} $ which assigns a polarity $+1$ to
  every move played by Eve (Player) and $-1$ to every move played by Adam
  (Opponent);
  of a prefix-closed set $\Plays{A}$ of finite paths, called the plays of the
  game~$A$.
  One requires moreover that every play of the game
  \[
    x_1\stackrel{e_1}{\longrightarrow}x_2 \stackrel{e_2}{\longrightarrow} \cdots
    \longrightarrow x_n \stackrel{e_n}{\longrightarrow} x_{n+1}
  \]
  is alternating in the sense that $\Polarity{A}(e_i)= (-1)^i$ for $1\leq i\leq
  n$, and that it starts and stops with an Adam move.
\end{defin}
A vertex in a game~$A$ is called \emph{initial} when there exists a play $s\in
\Plays{A}$ with $x=\partial_0(s)$ as source.
The set of initial vertices of a game~$A$ is noted~$\Initial{A}$.
We take below the most general and liberal definition of a strategy. In
particular, a strategy in that sense does not need to be deterministic.

\begin{defin}[Strategy]
  A strategy of a game is a prefix-closed set of plays.
  \end{defin}
  Every execution trace $t\in\Traces$ induces a game defined below, called the
  \emph{separation game} associated to $t$ and noted $\SeparationGame(t)$.
  \begin{defin}[Separation Game]
    The game $\SeparationGame(t)=(\Board{},\Polarity{},\Plays{})$ is defined as
    the graph $\Board{}=\GraphofSeparatedStates$ with plays in $\Plays{}$
    defined as the paths
    \[
      \play \quad : \quad (\state_C, \statevector,
      \state_F)\stackrel{*}{\longrightarrow} (\state'_C, \statevector',
      \state'_F)
    \]
    in $\GraphofSeparatedStates$ which combine into a path in $\GraphofStates$
\[\mbox{$\bigsprod$} \play  \quad : \quad
  \mbox{$\bigsprod$}(\state_C, \statevector,
  \state_F)\stackrel{*}{\longrightarrow} \mbox{$\bigsprod$}(\state'_C,
  \statevector', \state'_F)
\]
prefix of the trace~$t\in\Traces$.
The polarity $\Polarity{}$ of the moves is derived from the polarity Eve ($+1$)
and Adam ($-1$) of the edges of the graph $\Board{}=\GraphofSeparatedStates$ of
separated states.
\end{defin}
A play of the separation game $\SeparationGame(t)$ may be thus seen as a
``psychoanalysis'' or rather a ``couple therapy'' where Eve and Adam try and
justify \emph{a posteriori} what has just happened in the execution trace
$t\in\Traces$ played by the Code (on the side of Eve) and the Environment (on
the side of Adam).
At each transition $m:(\sigma,L)\to(\sigma',L')$ performed by the Code in the
execution trace~$t\in\Traces$ starting from a machine state
$(\sigma,L)=\bigsprod(\state_C, \statevector, \state_F)$, Eve has to play a move
$m:(\state_C, \statevector, \state_F)\to (\state'_C,\statevector',\state'_F)$
which ``justifies'' the transition by decomposing the machine state
$(\sigma',L')$ into a separated state~$(\state'_C, \statevector', \state'_F)$.
And symmetrically for Adam and the Environment.

\section{Soundness theorem}\label{sec:soundness}
At this stage, we establish our soundness theorem for concurrent separation
logic, by interpreting every derivation tree as a winning strategy in a specific
separation game.
We suppose given a Hoare triple $\triplestd$.
We start by describing the winning condition on the separation game
$\SeparationGame(t)$ associated to an execution trace $t\in\sem{C}$ in the
operational semantics of~$C$.

%
\begin{defin}
A \emph{separated predicate} is a triple $\mathbf{P}\, = \, (P, \Gamma, Q)$ consisting of two predicates $P$ and $Q$ and of a context $\Gamma=r_1:J_1,\dots,r_k:J_k$ of variable resources.
\end{defin}
\begin{defin}
We write
$$(\state_C, \statevector, \s_F) \vDash (P,\Gamma,Q)$$
and say that the separated state $(\state_C, \statevector, \s_F)$ satisfies the
separated predicate $\mathbf{P}=(P,\Gamma,Q)$ precisely when $\s_C \vDash P$ and
$\s_F \vDash Q$ and $\forall r \in \dom(\statevector), \statevector(r) \vDash
\D(r)$.
\end{defin}
%
We suppose from now on that the execution trace $t\in\sem{C}$ is of length $p$,
and introduce the sequence
%
%
$\mathbf{P}_1, \dots , \mathbf{P}_{2p+2}$ 
of separated predicates, defined as:
\[
\mathbf{P}_1 = (P,\G,\True)
\quad\quad
\mathbf{P}_i = (\True,\G,\True)
\quad\quad
\mathbf{P}_{2p+2} = (Q,\G,\True)
\]
for $1<i< 2p+1$ when the execution trace $t\in\returntraces{\sem{C}}$ is
returning; and defined as
\[
\mathbf{P}_1 = (P,\G,\True)
\quad\quad
\mathbf{P}_i = (\True,\G,\True)
\quad\quad
\mathbf{P}_{2p+2} = (\True,\G,\True)
\]
for $1<i< 2p+2$ when the execution trace $t\not\in\returntraces{\sem{C}}$ is not
returning.
Here, we write $\True$ for the constant predicate which is true for every
logical state.
%

\begin{defin}[Winning condition]\label{def/winning-play}
A play 
   \[(\state^1_C, \statevector^1, \state^1_F) \xrightarrow{\:env\:} 
   (\state^2_C, \statevector^2, \state^2_F) \xrightarrow{\:m_1\:}
   (\state^{3}_C, \statevector^{3}, \state^{3}_F)
   \xrightarrow{\;\;} \cdots \xrightarrow{\;\;}
   (\state^{2q+2}_C, \statevector^{2q+2}, \state^{2q+2}_F)
   \]
in the separation game $\SeparationGame(t)$
is declared winning when
\[
\forall i\in\{1,\ldots,2q+2\},\quad\quad\quad (\state^i_C, \statevector^i, \state^i_F)\vDash \mathbf{P}_i.
\]
\end{defin}
 %
Note that the notion of winning play is closed under prefix.
 
\begin{defin}
  A strategy $\textsf{strat}$ of the separation game $\SeparationGame(t)$ is
  winning when it contains only winning plays, and moreover:
  \begin{itemize}
  \item the strategy $\mathsf{strat}$ contains every empty and winning play of the separation game,
  \item for every play $\play$ in the strategy $\textsf{strat}$, which can be
    extended by a move $a$ played by Adam into a winning play $\play\cdot a$ of
    the separation game $\SeparationGame(t)$, there exists a move $e$ played by
    Eve such that $\play\cdot a\cdot e$ defines a play in the strategy
    $\textsf{strat}$.
  \end{itemize}
\end{defin}
Note that an empty and winning play of the separation game consists of a
separated state $(\state_C, \statevector, \state_F)$ satisfying the predicate
$(P,\Gamma,\True)$, and in the very special case when the trace
$t\in\returntraces{\sem{C}}$ is empty and returns, the predicate
$(Q,\Gamma,\True)$.

%
We are now able to state the soundness theorem of concurrent separation logic,
which is established by structural induction on the derivation tree $\pi$ of the
Hoare triple $\triplestd$.

\begin{thm}[Soundness]\label{thm:soundness}
  Every derivation tree $\pi$ of $\triplestd$ defines for every execution trace
  $t \in \sem{C}$ a winning strategy $\strat{\pi}{t}$ in the separation game
  $\SeparationGame(t)$ determined by the Hoare triple $\triplestd$ and $t$.
\end{thm}
The proof of the theorem is in the Appendix\cite{Appendix}.
This statement is inspired by game semantics, and the idea of a Curry-Howard
correspondence between proofs (derivation trees) and winning
strategies.
This interpretation of proofs implies the soundness of
concurrent separation logic in the traditional sense
\cite{Brookes:a-semantics,Vafeiadis,mezzo} by considering the case when the
context $\Gamma$ is empty, and the environment is passive, in the following
sense.

\begin{defin}
  The environment is passive in a trace
  \[
    \mstate_1\xrightarrow{env} \mstate_2 \xrightarrow{m_1}
    \mstate_3\xrightarrow{env} \ldots\xrightarrow{env}
    \mstate_{2p}\xrightarrow{m_p}\mstate_{2p+1}\xrightarrow{env}\mstate_{2p+2}
  \]
  when every transition $\mstate_{2i+1}\xrightarrow{env}\mstate_{2i+2}$ by the
  environment does not alter the logical state, and is thus the identity
  $\mstate_{2i+1}=\mstate_{2i+2}$, for $0\leq i \leq p$.
\end{defin}
\begin{cor}\label{cor:extensional-soundness}
  Suppose that the triple $\triple{\emptyset}{P}{C}{Q}$ has been proved by a
  derivation tree~$\pi$ of concurrent separation logic, and that $t \in \sem{C}$
  is an execution trace
    \[
    \mstate_1\xrightarrow{id} \mstate_1 \xrightarrow{m_1} 
    \mstate_3\xrightarrow{id}
    \ldots\xrightarrow{id}
    \mstate_{2p}\xrightarrow{m_p}\mstate_{2p+1}\xrightarrow{id}\mstate_{2p+1}
  \]
  in which the Environment is passive, and such that $\mstate_1 \vDash P*\True$.
  Then, the execution trace $t$ produces no error, in the technical sense that
  every machine step $m_i:\redto{s_{2i+1}}{}{s_{2i+3}}$ executed by the Code,
  for $0\leq i\leq p-1$ is of the form $\returnto{s_{2i+1}}{m_i}{s_{2i+3}}$ and
  thus does not produce any error at run-time. Moreover,
  when~$t\in\returntraces{C}$ returns, one has that:
  \[
    \partial_1t \vDash Q * \True.
  \]
\end{cor}
Note that the predicate $P\ast\True$ means that the logical state $\sigma$ taken
as input by the Code~$C$ contains a fragment $\sigma_C$ which satisfies the
predicate~$P$.
The winning strategy associated to $\pi$ ensures that when the trace $t$
returns, the Code~$C$ ends with a fragment $\sigma'_C$ of the logical state
$\sigma'$ returned as output.

\bibliographystyle{entcs}
\bibliography{citations}


\end{document}
